\documentclass[aps,prb,reprint,superscriptaddress,showpacs,citeautoscript]{revtex4-1}

\usepackage{graphicx}
\usepackage{color}

\begin{document}


\title{Giant Thermal Vibrations in the Framework Compounds Ba$_{1-x}$Sr$_x$Al$_2$O$_4$}


\author{S. Kawaguchi}
\email{kawaguchi@spring8.or.jp}
\affiliation{Japan Synchrotron Radiation Research Institute (JASRI/SPring-8), Sayo, Hyogo 679-5198, Japan.}

\author{Y. Ishii}
\email{ishii@mtr.osakafu-u.ac.jp}
\affiliation{Department of Materials Science, Osaka Prefecture University, Sakai, Osaka 599-8531, Japan.}

\author{E. Tanaka}
\affiliation{Department of Materials Science, Osaka Prefecture University, Sakai, Osaka 599-8531, Japan.}

\author{H. Tsukasaki}
\affiliation{Department of Materials Science, Osaka Prefecture University, Sakai, Osaka 599-8531, Japan.}

\author{Y. Kubota}
\affiliation{Department of Physical Science, Osaka Prefecture University, Sakai, Osaka 599-8531, Japan.}

\author{S. Mori}
\affiliation{Department of Materials Science, Osaka Prefecture University, Sakai, Osaka 599-8531, Japan.}



\date{\today}

\begin{abstract}
Synchrotron X-ray diffraction (XRD) experiments were performed on the network compounds Ba$_{1-x}$Sr$_x$Al$_2$O$_4$ at temperatures between 15 and 800 K.
The ferroelectric phase of the parent BaAl$_2$O$_4$ is largely suppressed by the Sr-substitution and disappears for $x\geq0.1$.
Structural refinements reveal that the isotropic atomic displacement parameter ($B_{\rm iso}$) in the bridging oxygen atom for $x\geq0.05$ is largely independent of temperature and retains an anomalously large value in the adjacent paraelectric phase even at the lowest temperature.
The $B_{\rm iso}$ systematically increases as $x$ increases, exhibiting an especially large value for $x\geq0.5$.
According to previous electron diffraction experiments for Ba$_{1-x}$Sr$_x$Al$_2$O$_4$ with $x\geq0.1$, strong thermal diffuse scattering occurs at two reciprocal points relating to two distinct soft modes at the M- and K-points over a wide range of temperatures below 800 K [Y. Ishii $et$ $al.$, Sci. Rep. \bf{6}\rm, 19154 (2016)].
Although the latter mode disappears at approximately 200 K, the former does not condense, at least down to 100 K.
The anomalously large $B_{\rm iso}$ observed in this study is ascribed to these soft modes existing in a wide temperature range.
\end{abstract}

\pacs{77.80.B-, 63.20.-e, 61.05.C-}

\maketitle


\section{Introduction}

Since the discovery of the incipient perovskite-type quantum paraelectric oxides\cite{QP-STO,QP-KTO}, the quantum criticality in ferroelectrics has attracted substantial interest in condensed-matter physics. 
On the border of ferroelectricity, the dielectric constant is enhanced in the vicinity of absolute zero.
This non-classical behavior was recently quantitatively explained\cite{QCFerro-NatPhys}, and the relevance of the ferroelectric quantum criticality to other condensed states, such as superconductivity in doped SrTiO$_3$\cite{STO-super1,STO-super2}, has been reported.
The recently observed quantum critical behavior in other new ferroelectric compounds, $e.g.$, the organic ferroelectric TSCaCl$_{2(1-x)}$Br$_{2x}$\cite{OrgFerro} and TTF-QBr$_{4-n}$I$_n$ complexes\cite{QP-TTF} and the multiferroic Ba$_2$CoGe$_2$O$_7$\cite{Ba2CoGe2O7}, has prompted fascinating studies on the quantum criticality in ferroelectrics.

\begin{figure}[t]
\includegraphics[width=88mm]{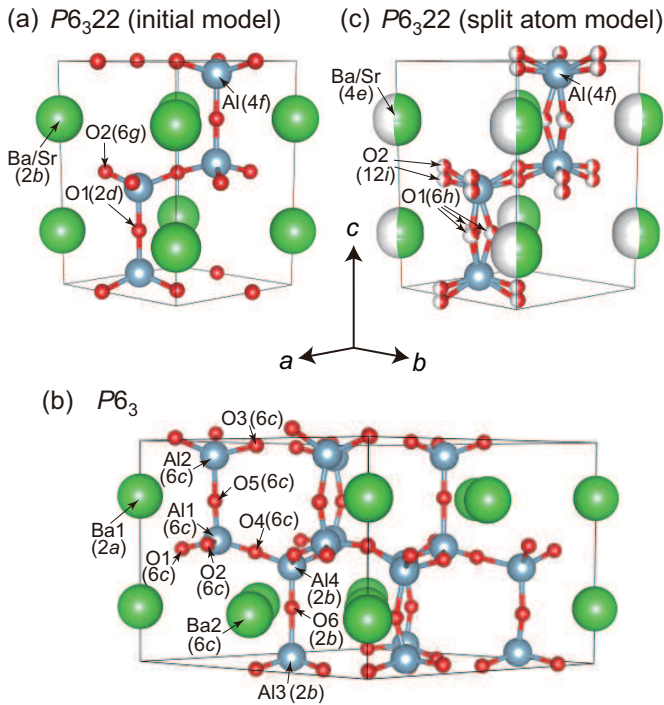}
\caption{\label{crystal_str} 
(Color online) Crystal structures of BaAl$_2$O$_4$. (a) The high-temperature phase (space group $P6_322$) and (b) the low-temperature phase ($P6_3$). (c) The split atom model for the $P6_322$ high-temperature phase. In this structure model, the atomic positions of Ba/Sr and O2 are split into two sites along the $c$-axis, and the O1 site is split into three sites around the threefold axis.}
\end{figure}

BaAl$_2$O$_4$ is a chiral improper ferroelectric\cite{Stokes} without an inversion center that  crystallizes in a stuffed tridymite-type structure comprising a corner-sharing AlO$_4$ tetrahedral network with six-member cavities occupied by Ba ions. 
This compound undergoes a structural phase transition from a high-temperature phase with a space group of $P6_322$ to a low-temperature phase with $P6_3$ at approximately 450 K\cite{Ishii_BaAl2O4,Huang1,Huang2}. This transition is accompanied by the significant tilting of the Al-O-Al bond angle along the $c$-axis, giving rise to the enlargement of the cell volume to $2a\times2b\times c$.
The crystal structures of the high-temperature and low-temperature phases are shown in Figs. \ref{crystal_str}(a) and (b), respectively.

This compound possesses low-energy phonon modes, which are associated with a tilting of the AlO$_4$ tetrahedra around the shared vertices without a large distortion in each AlO$_4$ block\cite{Perez-Mato}. 
Such low-energy phonon modes have also been reported in SiO$_2$ modifications\cite{RUMs1,RUMs2}, nepheline\cite{nepheline}, and ZrW$_2$O$_8$\cite{ZrW2O8-1} and are often called rigid unit modes (RUMs). 
RUMs can sometimes act as soft modes and cause structural phase transitions, as observed in quartz\cite{quartz2}, tridymite\cite{Pryde}, and nepheline\cite{nepheline}. RUMs have also been suggested as the origin of a negative thermal expansion, as in ZrW$_2$O$_8$\cite{ZrW2O8-1, ZrW2O8-2}.

Low-energy phonon modes, such as RUMs, can be observed as thermal diffuse scattering in electron and X-ray diffractions (XRD).
In electron diffraction experiments of BaAl$_2$O$_4$, a characteristic honeycomb-type diffuse scattering pattern has been reported over a wide range of temperatures below 800 K\cite{Abakumov,Ishii_BSAO}.
A similar pattern was also observed in Ba$_{1-x}$Sr$_x$Al$_2$O$_4$ with $x=0.4$\cite{Fukuda}.
In the structural refinements of the $x=0.4$ sample, significant disorder was noted in the two oxygen sites, O1($2d$) and O2($6g$), in the initial model of $P6_322$ symmetry, as shown in Fig \ref{crystal_str}(a). 
In those analyses, a split atom model, as shown in Fig. \ref{crystal_str}(c), was employed, in which the positions of Ba/Sr($2b$), O1($2d$), and O2($6g$) are off-center from their ideal positions and split into the less symmetric $4e$, $6h$ and $12i$ sites, respectively.
This type of disorder has also been reported in Sr$_{0.864}$Eu$_{0.136}$Al$_2$O$_4$\cite{HYamada} and BaGa$_2$O$_4$\cite{BaGa2O4}.

We recently investigated the low-energy phonon modes in BaAl$_2$O$_4$ in detail via synchrotron XRD using single crystals and first principles calculations\cite{Ishii_BaAl2O4}. 
According to the calculations, this compound possesses two unstable phonon modes at the M- and K-points with nearly the same energies.
Both of these unstable phonon modes gave rise to strong diffuse scattering intensities in a wide range of temperatures below 800 K.
Interestingly, their intensities sharply increase towards $T_{\rm C}$, indicating that the two modes soften simultaneously.

Furthermore, the ordered phase with the $P6_3$ superstructure has been reported to be substantially suppressed by a small amount of Sr-substitution for Ba\cite{Ishii_BSAO,Rodehorst}.
According to our electron diffraction experiments on Ba$_{1-x}$Sr$_x$Al$_2$O$_4$ with precisely controlled Sr concentrations, no superstructure is observed, at least down to 100 K, for $x\geq0.1$.
In the temperature and compositional window of $T<200$ K and $x\geq0.1$, although the soft mode at the K-point disappears, the soft mode at the M-point survives and shows further fluctuation as the temperature decreases\cite{Ishii_BSAO}. 
These findings imply the presence of a new quantum critical state induced by the soft modes in Ba$_{1-x}$Sr$_x$Al$_2$O$_4$.

In the present study, we performed synchrotron powder XRD on Ba$_{1-x}$Sr$_x$Al$_2$O$_4$ at 15--800 K, revealing an unusually large and temperature-independent thermal vibration at the bridging oxygens.



\section{Experimental}
Polycrystalline samples of Ba$_{1-x}$Sr$_x$Al$_2$O$_4$ ($x$ = 0, 0.02, 0.05, 0.06, 0.1, 0.3, and 0.5) were synthesized using a conventional solid-state reaction.
The sample preparation procedure is described elsewhere\cite{Ishii_BSAO,Tanaka}.
The obtained samples were stored in a vacuum.
The synchrotron powder XRD patterns were obtained in the temperature range of 15--800 K at the BL02B2 beamline of SPring-8\cite{02B2}.
The samples to be measured were crushed into fine powder and filled into a fused quartz capillary with a diameter of 0.2 mm. 
The diffraction intensities were recorded using an imaging plate and multiple microstrip solid-state detectors. 
The incident X-ray beam was monochromatized to 25 keV using a Si (111) double-crystal.
The temperature was controlled with flowing helium and nitrogen gases. 

The structure refinements were performed via the Rietveld method using the JANA2006 software package\cite{JANA}.
The split atom model was employed for the structural refinements of the high-temperature phase. For the fitting of several profiles obtained at high temperatures, the Ba/Sr atom was placed on the $2b$ site of the $P6_322$ average structure rather than on the $4e$ site of the split atom model because the atomic displacement from the $2b$ site is so small that the $4e$ site cannot be distinguished from the $2b$ site at these temperatures. The $P6_3$ structure model was used for the low-temperature phase below $T_{\rm C}$. Several profiles just below $T_{\rm C}$ were analyzed using the $P6_322$ split atom model because of the poor fitting results obtained using the $P6_3$ structure model. The space groups and the atomic positions used for the refinements are summarized in Table S1. 

\section{Results and Discussion}
\begin{figure}[t]
\includegraphics[width=88mm]{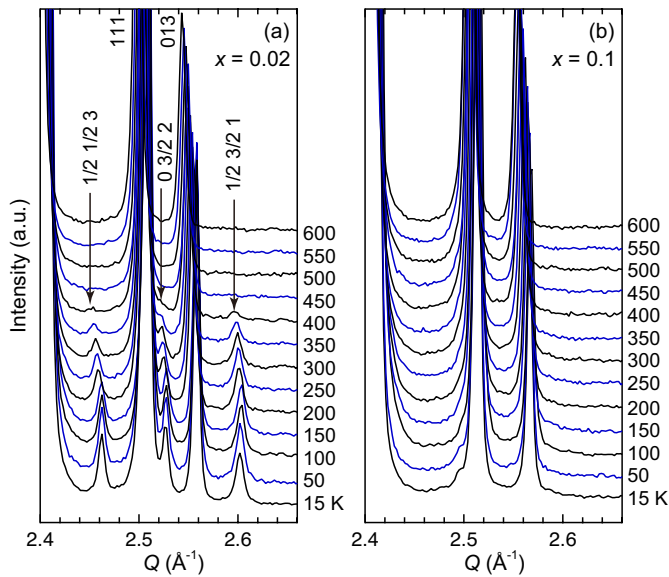}
\caption{\label{SXRD_Profiles} 
(Color online) Synchrotron powder XRD profiles obtained at 15--600 K for Ba$_{1-x}$Sr$_x$Al$_2$O$_4$ with (a) $x = 0.02$ and (b) 0.1. All of the indices throughout the paper are based on the $P6_322$ parent phase. The superlattice reflection grows below 400 K for $x=0.02$, as marked by the arrows, indicating the structural phase transition from $P6_322$ to $P6_3$. No superlattice reflections can be seen for the $x=0.1$ sample down to 15 K.}
\end{figure}

The obtained diffraction profiles for Ba$_{1-x}$Sr$_x$Al$_2$O$_4$ with $x=0.02$ and 0.1 are shown in Figs. \ref{SXRD_Profiles}(a) and (b), respectively, and 
the profiles for the other compositions are displayed in Fig. S1. 
The superlattice reflections of the low-temperature $P6_3$ phase can be clearly seen below 400 K for $x=0.02$, as indicated by arrows in Fig. \ref{SXRD_Profiles}(a).
These superlattice reflections were also observed for $x=0.05$ and 0.06, but they could not be observed for $x\geq0.1$.
Thus, the structural phase transition does not occur down to 15 K for $x\geq0.1$.

\begin{figure}[t]
\includegraphics[width=64mm]{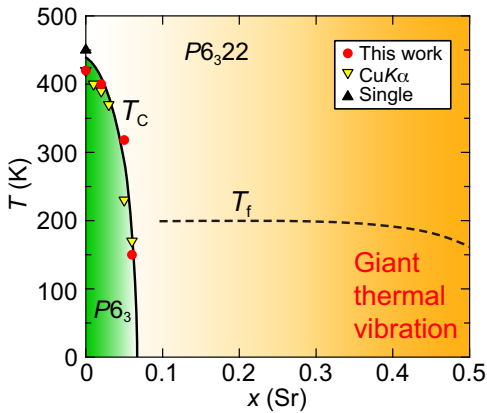}
\caption{\label{PhaseDiagram} 
(Color online) Phase diagram for Ba$_{1-x}$Sr$_x$Al$_2$O$_4$. $T_{\rm C}$ abruptly decreases as $x$ increases. 
Anomalously large and temperature-independent $B_{\rm iso}$ values were observed in the region of $x\geq0.1$, which is labeled as a giant thermal vibration.
$T_{\rm f}$ denotes the temperature at which the K-point mode disappears, as described in the text.}
\end{figure}

$T_{\rm C}$ was defined as the temperature at which the superlattice reflections appear.
A phase diagram for Ba$_{1-x}$Sr$_x$Al$_2$O$_4$ is shown in Fig. \ref{PhaseDiagram}.
The $T_{\rm C}$ values determined in this study were plotted together with the previously reported data\cite{Ishii_BSAO, Ishii_BaAl2O4}.
As shown in this figure, $T_{\rm C}$ becomes largely suppressed as $x$ increases, in agreement with previous reports\cite{Rodehorst,Ishii_BSAO}.
Synchrotron XRD experiments using single crystals\cite{Ishii_BaAl2O4} revealed a continuous variation of the superlattice intensities at $T_{\rm C}$, indicating a second-order phase transition.
These findings indicate that Ba$_{1-x}$Sr$_x$Al$_2$O$_4$ may plausibly show a new quantum critical state.

\begin{figure}[t]
\includegraphics[width=88mm]{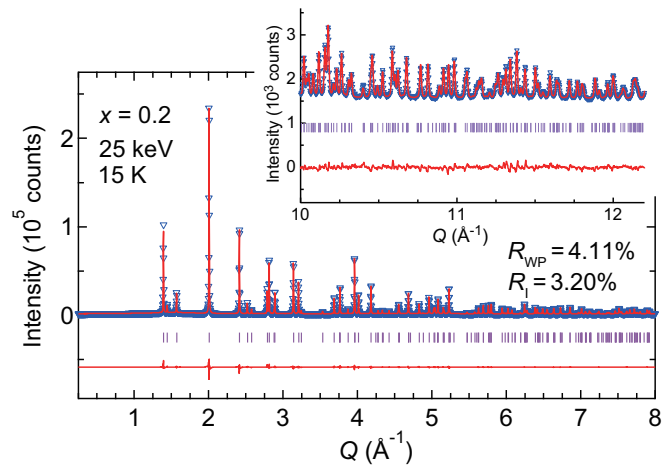}
\caption{\label{Rietveld} 
(Color online) Synchrotron powder XRD patterns (triangles) of Ba$_{0.8}$Sr$_{0.2}$Al$_2$O$_4$ at 15 K. The solid red curve represents the result of the structure refinement using the split atom model. The difference curve is shown in the lower part of the figure. Vertical lines indicate the positions of possible Bragg peaks. 
The fitting result for the high-$Q$ region is depicted in the inset. 
The Ba/Sr atom is placed at the $4e$ site.
}
\end{figure}

We performed structural refinements on the Ba$_{1-x}$Sr$_x$Al$_2$O$_4$ sample using the Rietveld method. 
Fig. \ref{Rietveld} displays the powder diffraction pattern for $x=0.2$ at 15 K and the fitting results obtained using the split atom model. 
The refinement using the initial model yielded a poor fitting result; the obtained $R$-factors based on the weighted profile ($R_{\rm WP}$) and the Bragg-integrated intensities ($R_{\rm I}$) were 5.14  and 9.25 \%, respectively.
In contrast, Fig. \ref{Rietveld} shows that the split atom model accurately reproduces the experimental profile with high reliability ($R_{\rm WP}$ = 4.11 \% and $R_{\rm I}$ = 3.20 \%).
The refined structural parameters are listed in Table \ref{Models}.
One important finding is that the isotropic atomic displacement parameter ($B_{\rm iso}$) of the O1 site is extraordinarily large, which is rare in oxide insulators.
Because the $R$-factors are sufficiently small, this large $B_{\rm iso}$ should have an important physical meaning.
Such a large $B_{\rm iso}$ in the O1 site has also been reported in previous works on Ba$_{0.6}$Sr$_{0.4}$Al$_2$O$_4$\cite{Fukuda}, for which it was ascribed to the thermal diffuse scattering observed in the diffraction experiments, and Sr$_{0.864}$Eu$_{0.136}$Al$_2$O$_4$\cite{HYamada} at room temperature.

 \begin{table}[t]
 \caption{\label{Models} Results of the structure refinement for Ba$_{0.8}$Sr$_{0.2}$Al$_2$O$_4$ at 15 K using the split atom model. The cation ratio of Ba:Sr is fixed to 0.8:0.2. 
}
 \begin{ruledtabular}
 \begin{tabular}{ccccccc}
 Atom & site & $g$ & $x$ & $y$ & $z$ & $B_{\rm iso}$ (\AA$^{2}$) \\ \hline
 Ba/Sr & 4$e$ & 1/2 & 0 & 0 & 0.2414(1) & 0.392(7) \\ 
 Al & 4$f$ & 1 & 1/3 & 2/3 & 0.9451(1) & 0.93(2) \\
 O1 & 6$h$ & 1/3 & 0.3863(3) & 2$x$ & 3/4 & 2.1(2) \\
 O2 & 12$i$ & 1/2 & 0.3593(8) & -0.002(1) & 0.0289(3) & 0.77(5) \\
 \end{tabular}
 \end{ruledtabular}
 \end{table}

\begin{figure}[t]
\includegraphics[width=87mm]{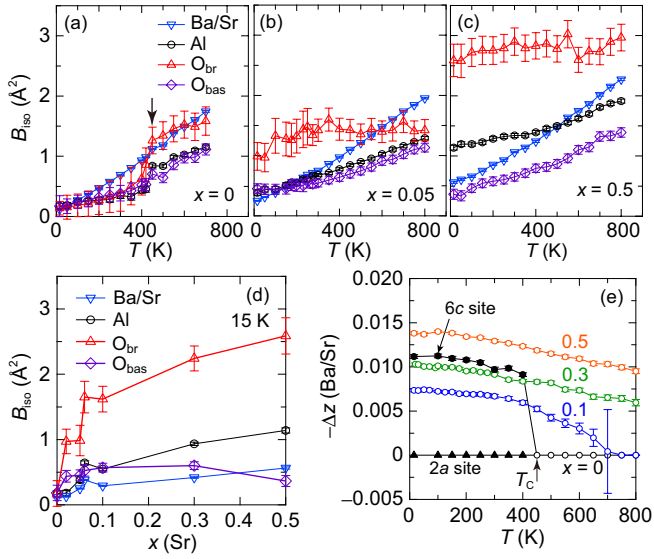}
\caption{\label{Biso} 
(Color online) Temperature dependence of $B_{\rm iso}$ of Ba/Sr, Al, the bridging oxygen (O$_{\rm br}$) and the basal oxygen (O$_{\rm bas}$) atoms for (a) $x = 0$, (b) 0.05, and (c) 0.5. $T_{\rm C}$ is indicated by an arrow. (d) $B_{\rm iso}$ at 15 K plotted as a function of $x$. (e) The displacement of the Ba/Sr atom ($-\Delta z$) from the ideal $2b$ site of $z=0.25$. 
The $-\Delta z$ values obtained using the $P6_322$ split atom model are indicated by the open symbols.
Below 420 K, the Ba site for $x=0$ is split into two sites of $P6_3$, $2a$ and $6c$, as indicated by the closed symbols.
}
\end{figure}

The temperature dependences of the obtained $B_{\rm iso}$ for the $x$ = 0, 0.05, and 0.5 samples are shown in Figs. \ref{Biso}(a), (b), and (c), respectively. 
In these figures, the bridging oxygen, O$_{\rm br}$, and the basal oxygen, O$_{\rm bas}$, represent the O1 and O2 atoms in the $P6_322$ structure, respectively. 
In the structural refinements for the $P6_3$ structure, the atoms were categorized into four groups: the Ba group of Ba1--2, the Al group of Al1--4, the basal oxygen group of O1--4, and the bridging oxygen group of O5--6.
The calculations were performed under the constraint that an equal $B_{\rm iso}$ is used within each group.
As shown in Fig. \ref{Biso}(a), the $B_{\rm iso}$ of each atom for $x=0$ gradually decreases as the temperature decreases, and rapidly decreases at $T_{\rm C}$, except for the Ba atom, as marked by an arrow in Fig. \ref{Biso}(a).
This is probably because of the structural phase transition accompanying the condensation of one of the soft modes and the resulting suppression of the thermal vibrations.
The $B_{\rm iso}$ values of all atoms are small at 15 K.
The sudden drops in the $B_{\rm iso}$ at $T_{\rm C}$ were also observed for the $x=0.02$ sample.
The temperature dependences of the $B_{\rm iso}$ for all compositions are shown in Fig. S2.

The full width at half maximum (FWHM) of the superlattice reflection has been reported to increase as $x$ increases\cite{Ishii_BSAO}, as observed in this study (Fig. S3).
Thus, the long-range ordering of the low-temperature $P6_3$ structure is strongly suppressed by the Sr-substitution.
In Fig. \ref{Biso}(b), no sudden drop in $B_{\rm iso}$ can be observed for $x=0.05$, although the superlattice reflections develop below 320 K in the $x = 0.05$ sample, as shown in Fig. S1(c).
This is probably because of the suppression of the long-range ordering of the low-temperature structure in the $x=0.05$ sample.

In contrast, the observed $B_{\rm iso}$ for $x=0.5$ is surprisingly large, as seen in Fig. \ref{Biso}(c).
In addition, it is largely independent of the temperature, resulting in anomalously large values even at the lowest temperature. 
Such a temperature-independent large $B_{\rm iso}$ is also observed for $x=0.1$ and 0.3, as shown in Figs. S2(e) and (f).
Other atoms also exhibit larger $B_{\rm iso}$ over the whole temperature range than that of $x=0$.
Because the $B_{\rm iso}$ values for Ba/Sr, Al, and O$_{\rm bas}$ are sufficiently small at 15 K, and the refinements yield the satisfactorily small $R$-factors, as shown in Figs. S4 and S5, the extraordinarily large $B_{\rm iso}$ observed in O$_{\rm br}$ cannot be attributed to the fitting error in the refinements.
Fig. \ref{Biso}(d) represents the $B_{\rm iso}$ for each atom at 15 K as a function of $x$.
These values systematically increase as $x$ increses, particularly for the O$_{\rm br}$ atom.
The $B_{\rm iso}$ for the O$_{\rm br}$ atoms are especially large for $x\geq0.06$, indicating that the O$_{\rm br}$ atom for $x\geq0.06$ exhibits an unusually large thermal vibration down to absolute zero.
Notably, this enhancement in $B_{\rm iso}$ is observed on the border of the ferroelectric phase.

The Ba/Sr atom displaces only along the $c$-axis in the split atom model.
The displacements of the Ba/Sr atoms ($-\Delta z$) from the ideal $2b$ site of $z=0.25$ are plotted in Fig. \ref{Biso}(e) as a function of temperature. 
These gradually increase as the temperature decreases. 
For $x=0$, the Ba site splits into the $2a$ and $6c$ sites of the $P6_3$ low-temperature structure at $T_{\rm C}$, as indicated by the closed symbols in Fig. \ref{Biso}(e), because of the structural phase transition.
Below $T_{\rm C}$, the $-\Delta z$ of Ba at the $6c$ site increases as the temperature decreases.
For $x=0.1$, 0.3, and 0.5, the $-\Delta z$ increases as $x$ increases.
Notably, the $-\Delta z$ for $x=0.5$ exceeds that for $x=0$ in the ferroelectric phase, although the $x=0.5$ sample does not exhibit a structural phase transition.

The temperature and compositional window of the anomalously large $B_{\rm iso}$ is illustrated in Fig. \ref{PhaseDiagram}. 
In general, the thermal vibration is large at high temperatures and becomes suppressed as temperature decreases, and thus $B_{\rm iso}$ is expected to decrease as the temperature drops.
However, the $B_{\rm iso}$ of the O$_{\rm br}$ atom in this system with $x\geq0.1$ does not follow this general trend; instead, it exhibits fairly large values even at 15 K. 
According to our previous studies, this system possesses  structural instabilities at the M- and K-points, leading to two energetically competing soft modes\cite{Ishii_BaAl2O4}.
Mode analyses have revealed that the O$_{\rm br}$ atom shows a remarkable in-plane vibration around the ideal $2d$ site.
In addition, strong thermal diffuse scattering because of these soft modes has been observed in the electron diffraction patterns for $x\geq0.1$ over a wide range of temperatures between 100 and 800 K\cite{Ishii_BSAO}.
Clearly, these soft modes are responsible for the unusually large $B_{\rm iso}$ observed in this study.
The large enhancement in $B_{\rm iso}$ by the Sr substitution indicates ``a giant thermal vibration,'' which might be a phonon-related quantum critical phenomenon.
For the $x\geq0.1$ sample, the M-point soft mode does not condense but survives down to at least 100 K, whereas the K-point mode weakens and disappears below 200 K\cite{Ishii_BSAO}.
The temperature at which the K-point mode disappears, $T_{\rm f}$, is indicated by a broken line in Fig. \ref{PhaseDiagram}.
Below this line, the M-point mode fluctuates with short-range correlations in nanoscale regions\cite{Ishii_BaAl2O4}.
The anomalously large and temperature-independent $B_{\rm iso}$ even at low temperatures can be attributed to this fluctuating M-point mode.

Large thermal displacements have also been reported in the metallic phase of VO$_2$\cite{VO2-1, VO2-2}, which exhibits a drastic metal-insulator transition (MIT).
Several approaches, including thermal diffuse scattering just above the MIT\cite{VO2-TDS}, the symmetry analyses\cite{VO2-Group}, and the phonon dispersion calculations\cite{VO2-cal}, indicate the presence of a soft mode.
ZrW$_2$O$_8$ is another example that shows a large $B_{\rm iso}$ in its oxygen atoms, comprises a corner-sharing polyhedral network, and possesses RUMs.
However, the $B_{\rm iso}$ values of these oxygen atoms have been reported to decrease as the temperature decreases\cite{ZrW2O8_Biso}.
No compound showing a temperature-independent and anomalously large thermal vibration even at low temperature has been reported to date.
To clarify the nature of the fluctuating state in Ba$_{1-x}$Sr$_x$Al$_2$O$_4$, measurements of the physical properties, such as the dielectric constant and specific heat, are now in progress.

\section{Conclusions}

We performed synchrotron XRD experiments for Ba$_{1-x}$Sr$_x$Al$_2$O$_4$ at 15--800 K and analyzed their crystal structures via the Rietveld method using the split atom model.
$T_{\rm C}$ becomes substantially suppressed as $x$ increases, and the structural phase transition disappears for $x\geq0.1$.
We observed an anomalously large and temperature-independent $B_{\rm iso}$ in the bridging oxygen for the $x\geq0.1$ samples, with the value systematically increasing as $x$ increases.
These anomalously large $B_{\rm iso}$ values can be attributed to the giant thermal vibration arising from the existence of soft modes over a wide temperature range.

\begin{acknowledgments}
This work was partially supported by a Grant-in-Aid for Scientific Research from the Ministry of Education, Culture, Sports, Science and Technology of Japan (MEXT) from the Japan Society for the Promotion of Science (JSPS).
The synchrotron radiation experiments were performed at BL02B2 of SPring-8 with the approval of the Japan Synchrotron Radiation Research Institute (JASRI) (Proposals No. 2015A2058, No. 2015A1510, and No. 2015B1488).
\end{acknowledgments}


\end{document}